\documentclass[aps,onecolumn,showpacs,showkeys,nofootinbib]{revtex4}
\usepackage{epsfig}
\usepackage{amsmath}
\usepackage{amsfonts}
\usepackage{amssymb}
\usepackage{graphicx}
\usepackage{colordvi}
\begin{document}

\title{Rotation Curves of Galaxies by Fourth Order Gravity}

\author{$^1$A. Stabile\footnote{e -
mail address: arturo.stabile@gmail.com}, G. Scelza\footnote{e -
mail address: lucasce73@gmail.com}}

\affiliation{$^1$Dipartimento di Ingegneria, Universita' del
Sannio, Palazzo Dell'Aquila Bosco Lucarelli, Corso Garibaldi, 107
- 82100, Benevento, Italy}

\begin{abstract}

We investigate the radial behavior of galactic rotation curves by a Fourth Order Gravity adding also the Dark Matter component. The Fourth Order Gravity is a theory of Gravity described by Lagrangian generalizing the one of Hilbert Einstein containing a generic function of the Ricci scalar, the Ricci and Riemann tensor. A systematic analysis of rotation curves, in the Newtonian Limit of theory, induced by all galactic sub-structures of ordinary matter is shown. This analysis is presented for Fourth Order Gravity with and without Dark Matter. The outcomes are compared with respect to the classical outcomes of General Relativity. The gravitational potential of point-like mass is the usual potential corrected by two Yukawa terms. The rotation curve is higher or also lower than curve of General Relativity if in the Lagrangian the Ricci scalar square is dominant or not with respect to the contribution of the Ricci tensor square. The theoretical spatial behaviors of rotation curve are compared with the experimental data for the Milky Way and the galaxy NGC 3198. Although the Fourth Order Gravity gives more rotational contributions, in the limit of large distances the Keplerian behavior is ever present, and it is missing only if we add the Dark Matter component. However by modifying the theory of Gravity, consequently, also the spatial description of Dark Matter could undergo a modification and the free parameters of model can assume different values. After an analytical discussion of theoretical behaviors and the comparing with experimental evidence we can claim that any Fourth Order Gravity is not successful to explain the galactic rotation curves. In the last part of paper we analyze the gravitational potential induced by Lagrangian containing only powers of Ricci scalar. In this case we find an inconsistency in the boundary conditions in the passage from matter to the vacuum.

\end{abstract}
\maketitle

\section{Introduction}

Today the Universe appears spatially flat undergoing an accelerated expansion. There are many measurements proving this pictures \cite{riess, ast, clo, ber, spe}. According to the successful cosmological model \cite{carrol, staro, sahini}, there are two main ingredients in this scenario, namely Dark Matter (DM) and the cosmological constant $\Lambda$ (Dark Energy). On the galactic scales, the evolution is driven by the usual Newtonian gravitational potential, but it needs hypothesizing the existence of DM to obtain a good experimental agreement. A good model for the galactic distribution of DM, in the framework of General Relativity (GR), is the Navarro-Frenk-White model (NFW model) \cite{navarro}.

However in recent years, the effort to give a physical explanation to the cosmic acceleration has attracted an amount of interest in so called Fourth Order Gravity (FOG), and particularly the $f(R)$-Gravity, where $f$ is a generic function of Ricci scalar $R$ and the lagrangian of the theory is $\mathcal{L}\,=\,f(R)$. These alternative models have been considered as a viable mechanism to explain the cosmic acceleration. Other issues, of astrophysical nature, as the observed Pioneer anomaly problem \cite{anderson1, anderson2} can be framed into the same approach \cite{bertolami}, even if the anomaly has been shown to be attributable to well-understood classical mechanisms of radiation reaction \cite{mur, kat}. Apart the cosmological dynamics, a systematic analysis of such theories were performed at short scale and in the low energy limit \cite{olmo1, olmo2, olmo3, Damour:Esposito-Farese:1992, clifton, odintsov, newtonian_limit_fR, PRD, stelle, schm, FOG, Stabile_Capozziello}.

Although FOG is investigated at theoretical level as alternatives to GR with Dark Energy and /or DM, an analytical approach to galactic rotation curves have been considered for $f(R)$-Gravity \cite{Boehmer:2007kx, Boehmer:2007fh} and for a $R^n$-Gravity \cite{CCCT, CCT}, with $n\,\epsilon\,\mathrm{Q}$, where the Hilbert-Einstein Lagrangian, $\mathcal{L}\,=\,R$, is substituted by $\mathcal{L}\,=\,R^n$. In fact in these papers one does not consider a correction to scalar Ricci but one substitutes it by a power law for the Ricci scalar. In such way one find yet a Newtonian gravitational potential ($\propto\,r^{-1}$) with an added power law correction.

By evaluating the rotation curve for a generic gravitational potential we want, in this paper, to use the potential induced by a $f(X,Y,Z)$-Gravity, where for sake of simplicity we set $X\,=\,R$, $Y\,=\,R^{\alpha\beta}R_{\alpha\beta}$ and $Z\,=\,R^{\alpha\beta\gamma\delta}R_{\alpha\beta\gamma\delta}$. Then we generalize the Hilbert Einstein lagrangian by introducing a generic function depending not only on the Ricci scalar $R$ but also on other two curvature invariants $Y$, $Z$ where $R_{\mu\nu}$ is the Ricci tensor
and $R_{\mu\nu\alpha\beta}$ is the Riemann tensor. The gravitational potential is obtained when the Newtonian limit of theory with lagrangian $\mathcal{L}\,=\,f(X,Y,Z)$ is performed \cite{FOG}. We are here interested to analyze the induced corrections when one considers the rotation curves. The corrections to potential can modify the global rotation curve and the DM model can undergo a modification. We report the radial behavior of rotation curve for two galaxies: Milky Way and NGC 3198. We analyze for any galactic substructure the rotation curve contribution in all frameworks: GR, GR with DM, FOG and FOG with DM.

Since, today, there is not a successful model to explain the rotation curve without requiring the existence of DM, but there are many partially acceptable model, we conclude this work by comparing the outcomes of this paper with respect to  the ones of $R^n$-Gravity. In this last topic we consider the analogy and the differences among the two theories by valuating also their metrics and Minkowskian limit.

The plan of the paper is as follows. In Section \ref{grav_pot}, we present the modified gravitational potential used. Section \ref{curve} describes the rotation curve and the main properties induced in the potential. In the sections \ref{mass_model} there is a little classification of principal galactic mass model and in section \ref{curve_FOG} we show the theoretical predictions of rotation curves and compare them with respect to the experimental data; section \ref{CCT_theory} resumes the analogies and differences between FOG and $R^n$-Gravity, while Section \ref{conclusion} summarizes our principal conclusions.

\section{The gravitational potential by $f(X,Y,Z)$-Gravity}\label{grav_pot}

Let us start with a general class of FOG given by the action

\begin{eqnarray}\label{FOGaction}
\mathcal{A}\,=\,\int d^{4}x\sqrt{-g}\biggl[\mathcal{L}+\mathcal{X}\mathcal{L}_m\biggr]\,=\,\int d^{4}x\sqrt{-g}\biggl[f(X,Y,Z)+\mathcal{X}\mathcal{L}_m\biggr]
\end{eqnarray}
where $f$ is an unspecified function of curvature invariants.
The term $\mathcal{L}_m$ is the minimally coupled ordinary matter
contribution. In the metric approach, the field equations are
obtained by varying (\ref{FOGaction}) with respect to
$g_{\mu\nu}$. We get

\begin{eqnarray}\label{fieldequationFOG}
\left\{\begin{array}{ll}
H_{\mu\nu}\,=\,f_XR_{\mu\nu}-\frac{f}{2}g_{\mu\nu}-f_{X;\mu\nu}+g_{\mu\nu}\Box
f_X+2f_Y{R_\mu}^\alpha
R_{\alpha\nu}-2[f_Y{R^\alpha}_{(\mu}]_{;\nu)\alpha}\\\\\,\,\,\,\,\,\,\,\,\,\,\,\,\,\,\,\,\,\,\,\,\,\,\,\,\,\,\,\,\,\,\,\,\,
\,\,\,\,\,\,\,\,\,\,\,\,\,\,\,+\Box[f_YR_{\mu\nu}]+[f_YR_{\alpha\beta}]^{;\alpha\beta}g_{\mu\nu}
+2f_ZR_{\mu\alpha
\beta\gamma}{R_{\nu}}^{\alpha\beta\gamma}-4[f_Z{{R_\mu}^{\alpha\beta}}_\nu]_{;\alpha\beta}\,=\,
\mathcal{X}\,T_{\mu\nu}
\\\\\\\\
H\,=\,f_XR+2f_YR_{\alpha\beta}R^{\alpha\beta}+2f_ZR_{\alpha\beta\gamma\delta}
R^{\alpha\beta\gamma\delta}-2f+\Box[3
f_X+f_YR]+2[(f_Y+2f_Z)R^{\alpha\beta}]_{;\alpha\beta}\,=\,\mathcal{X}\,T
\end{array}\right.
\end{eqnarray}
where $f_X\,=\,\frac{df}{dX}$, $f_Y\,=\,\frac{df}{dY}$,
$f_Z\,=\,\frac{df}{dZ}$, $\Box={{}_{;\sigma}}^{;\sigma}$ and
$\mathcal{X}\,=\,8\pi G$\footnote{Here we use the convention
$c\,=\,1$.}.
$T_{\mu\nu}\,=\,-\frac{1}{\sqrt{-g}}\frac{\delta(\sqrt{-g}\mathcal{L}_m)}{\delta
g^{\mu\nu}}$ is the energy-momentum tensor of matter and $T$ is
its trace. The second line of (\ref{fieldequationFOG}) is the
trace of the first one.

In the case of weak field and slow motion we consider the field
equation in the so called Newtonian limit of theory. For our aim
we can consider the metric tensor approximated as follows (for
details, see \cite{newtonian_limit_fR, rew, landau, PRD})

\begin{eqnarray}\label{metric_tensor_PPN}
  g_{\mu\nu}\,=\,\begin{pmatrix}
  1+2\,\Phi(t,\mathbf{x})& 0 \\
  \\
  0 & -[1-2\Psi(t,\mathbf{x})]\delta_{ij}\end{pmatrix}
\end{eqnarray}
where $\Phi$ and $\Psi$ are the gravitational potentials and $\delta_{ij}$ is
the Kronecker delta. The set of coordinates\footnote{The Greek
index runs from $0$ to $3$; the Latin index runs from $1$ to $3$.}
adopted is $x^\mu\,=\,(t,x^1,x^2,x^3)\,=\,(t,\mathbf{x})$. By
introducing the quantities

\begin{eqnarray}\label{mass_definition}
\left\{\begin{array}{ll}
{m_1}^2\,\doteq\,-\frac{f_X(0)}{3f_{XX}(0)+2f_Y(0)+2f_Z(0)}\\\\
{m_2}^2\,\doteq\,\frac{f_X(0)}{f_Y(0)+4f_Z(0)}
\end{array}\right.
\end{eqnarray}
we get three differential equations for the curvature invariant $X$
and the gravitational potentials $\Phi$, $\Psi$\footnote{Throughout the paper we assume always $f_X(0)\,>\,0$, and therefore we may set $f_X(0)\,=\,1$ without loss of generality.}

\begin{eqnarray}\label{NL-field-equation_2}
\left\{\begin{array}{ll}
(\triangle-{m_2}^2)\triangle\Phi+\biggl[\frac{{m_2}^2}{2}-\frac{{m_1}^2+2{m_2}^2}{6{m_1}^2}\triangle\biggr]
X\,=\,-{m_2}^2\mathcal{X}\,\rho\\\\
(\triangle-{m_2}^2)R_{ij}+\biggl[\frac{{m_1}^2-{m_2}^2}{3{m_1}^2}\,
\partial^2_{ij}-\biggl(\frac{{m_2}^2}{2}-\frac{{m_1}^2+2{m_2}^2}{6{m_1}^2}\triangle\biggr)\delta_{ij}\biggr]
X\,=\,0\\\\
(\triangle-{m_1}^2)X\,=\,{m_1}^2\mathcal{X}\,\rho
\end{array}\right.
\end{eqnarray}
where $\triangle$ is the Laplacian in the flat space, $R_{ij}\,=\,\triangle\Psi\,\delta_{ij}+(\Psi-\Phi)_{,ij}$ is the $ij$-component of Ricci tensor and $\rho$ is the matter density \cite{FOG}.

By choosing ${m_1}^2\,,{m_2}^2\,>0$ and introducing $\mu_{1,2}\,\doteq\,\sqrt{|{m_{1,2}}^2|}$ the gravitational
potentials in the case of point-like
source ($\rho\,=\,M\,\delta(\mathbf{x})$) are given by

\begin{eqnarray}\label{sol_pointlike}
\left\{\begin{array}{ll}
\Phi_{pl}(\mathbf{x})\,=\,-\,\frac{GM}{|\textbf{x}|}\biggl[1+\frac{1}{3}\,e^{-\mu_1|\mathbf{x}|}-\frac{4}{3}\,
e^{-\mu_2|\mathbf{x}|}\biggr]\\\\
\Psi_{pl}(\mathbf{x})\,=\,-\,\frac{GM}{|\textbf{x}|}\biggl[1-\frac{1}{3}\,e^{-\mu_1|\mathbf{x}|}-\frac{2}{3}\,
e^{-\mu_2|\mathbf{x}|}\biggr]
\end{array}\right.
\end{eqnarray}
If we have a generic matter source distribution $\rho(\mathbf{x})$, $\Phi_{pl}(\mathbf{x})$ becomes

\begin{eqnarray}\label{sol_gen}
\Phi(\mathbf{x})\,=\,-\,G\int
d^3\mathbf{x}'\frac{\rho(\mathbf{x}')}{|\mathbf{x}-\mathbf{x}'|}
\biggl[1+\frac{1}{3}\,e^{-\mu_1|\mathbf{x}-\mathbf{x}'|}-\frac{4}{3}\,e^{-\mu_2|\mathbf{x}-\mathbf{x}'|}\biggr]
\end{eqnarray}
and an analogous relation is found for $\Psi$. The solution (\ref{sol_gen}) has been obtained by using the
superposition principle by starting from the solution (\ref{sol_pointlike}). This approach is correct only in the Newtonian limit since a such limit correspond also to the linearized version of theory. The $f(X,Y,Z)$-Gravity (like GR) is not linear, then we would have had to solve the field equations (\ref{fieldequationFOG}) for a given matter density.

The parameters $\mu_i$ are the wave vectors but at same time in the fields theory are also the masses of propagation particles of field. In fact the equations (\ref{NL-field-equation_2}) are the Newtonian limit (\emph{i.e.} weak field limit and small velocity) of field equations (\ref{fieldequationFOG}), while if we perform only the weak field limit of (\ref{fieldequationFOG}) we obtain the propagation of Ricci scalar (trace equation) and one of Ricci tensor (tensorial field equation)

\begin{eqnarray}\label{wave-field-equation}
\left\{\begin{array}{ll}
(\Box+{m_2}^2)R_{\mu\nu}\,=\,source\\\\
(\Box+{m_1}^2)X\,=\,source
\end{array}\right.
\end{eqnarray}
where $\Box$ is the d'Alembert operator in the flat space. Then $\mu_1$ and $\mu_2$ are the masses linked respectively to the propagation of Ricci scalar and Ricci tensor.

If we choose the derivatives of $f$ with respect to the curvature invariants satisfying the condition $f_{XX}(0)+f_{Y}(0)+2f_{Z}(0)\,=\,0$ we find $\mu_1\,=\,\mu_2\,=\,\mu$ and the point-like solutions (\ref{sol_pointlike}) become

\begin{eqnarray}\label{sol_pointlike_A}
\Phi_{pl}(\mathbf{x})\,=\,\Psi_{pl}(\mathbf{x})\,=\,-\frac{GM}{|\textbf{x}|}\biggl[1-e^{-\mu|\mathbf{x}|}\biggr]
\end{eqnarray}
then it is verified the condition $g_{tt}\,g_{rr}\,\sim\,-1$. A such condition is satisfied by the spherically symmetric metrics (for example Schwarzschild, Schwarzschild - de Sitter, Einstein - de Sitter, Reissner - Nordostr\"{o}m, etc). In fact in the paper \cite{FOG} one has the condition $\triangle(\Phi_{pl}-\Psi_{pl})\,=\,0$ if $\mu_1\,=\,\mu_2$. Then we can affirm that only in GR the metric potentials $\Phi_{pl}$ and
$\Psi_{pl}$ are equals (or more generally their difference must be proportional to function $|\mathbf{x}|^{-1}$), while in FOG the Yukawa corrections must be equals.

\section{Rotation curves of galaxies}\label{curve}

The motion of body embedded in the gravitational field is given by geodesic equation

\begin{eqnarray}\label{geodesic}
\frac{d^2\,x^\mu}{ds^2}+\Gamma^\mu_{\alpha\beta}\frac{dx^\alpha}{ds}\frac{dx^\beta}{ds}\,=\,0
\end{eqnarray}
where $ds\,=\,\sqrt{g_{\alpha\beta}dx^\alpha dx^\beta}$ is the relativistic distance and $\Gamma^\mu_{\alpha\beta}$ are the Christoffel symbols. In the Newtonian limit of theory we obtain from (\ref{geodesic}), formally, the classical structure of the motion equation

\begin{eqnarray}
\frac{d^2\,\mathbf{x}}{dt^2}\,=\,-\nabla\Phi(\mathbf{x})
\end{eqnarray}
but the gravitational potential is given by (\ref{sol_gen}). A such potential is modified with respect to classical potential since we introduced other curvature invariants in the action (\ref{FOGaction}). If we want to come back in the theory we have to set $f_{XX}\,=\,f_Y\,=\,f_Z\,=\,0$, then $\mu_1\,,\mu_2\,\rightarrow\,\infty$ and the (\ref{sol_gen}) becomes the classical potential.

The study of motion is very simple if we consider a particular symmetry of mass distribution $\rho$, otherwise the analytical solutions are not available. Our aim is to evaluate the corrections to the classical motion in the easiest situation: the circular motion. In this case we do not consider the radial and vertical motion. The condition of stationary motion on the circular orbit is

\begin{eqnarray}\label{stazionary_motion}
\frac{{v_c(|\mathbf{x}|)}^2}{|\mathbf{x}|}\,=\,\frac{\partial\Phi(\mathbf{x})}{\partial|\mathbf{x}|}
\end{eqnarray}
where $v_c$ is the velocity.

The distribution of mass can be modeled simply by introducing two sets of coordinates: the spherical coordinates $(r,\theta,\phi)$ and the cylindrical coordinates $(R,\theta,z)$. An useful mathematical tool is the Gauss flux theorem for Gravity: \emph{The gravitational flux through any closed surface is proportional to the enclosed mass}. The law is expressed in terms of the gravitational field. The gravitational field $\mathbf{g}$ is defined so that the gravitational force experienced by a particle with mass $m$ is $\mathbf{F}_{grav}\,=\,m\,\mathbf{g}$. Since the Newtonian mechanics satisfies this theorem and, by thinking to a spherical system of mass distribution, we get, from (\ref{stazionary_motion}), the equation

\begin{eqnarray}\label{circular_velocity}
{v_c(r)}^2\,=\,\frac{G\,M(r)}{r}\,=\,\frac{4\pi G}{r}\int_0^rdy\,y^2\,\rho(y)
\end{eqnarray}
where $M(r)$ is the only mass enclosed in the sphere with radius $r$. The Green function of the $f(X,Y,Z)$-Gravity ($\neq\,|\mathbf{x}-\mathbf{x}'|^{-1}$), instead, does not satisfy the theorem \cite{Stabile_Capozziello}. In this case we must consider directly the gravitational potential (\ref{sol_gen}). Apart the mathematical difficulties incoming from the research of gravitational potential for a given mass distribution, the non-validity of Gauss theorem implies, for example, that a sphere can not be reduced to a point. In fact the gravitational potential generated by a ball (also with constant density) is depending also on the Fourier transform of ball \cite{Stabile_Capozziello}. Only in the limit case where the radius of ball is small with respect to the distance we obtain the simple expression (\ref{sol_pointlike}). However in this paper we want to consider not the simple case of motion of body in the vacuum, but the more interesting case of motion in the matter. So we must leave any possibility of idealization and consider directly the calculation of the potential (\ref{sol_gen}).

\begin{figure}[htbp]
  \centering
  \includegraphics[scale=1]{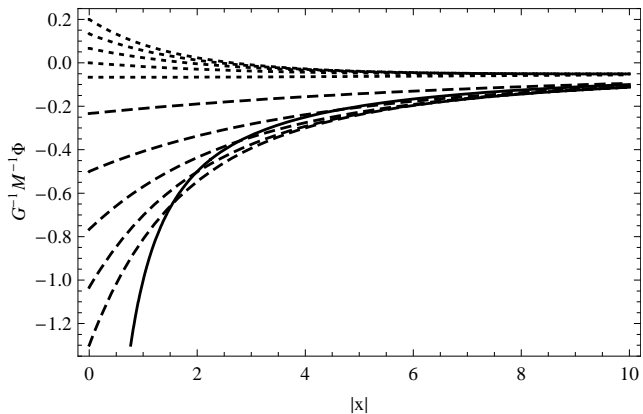}\\
  \caption{Plot of gravitational potential (\ref{sol_pointlike}). $\mu_2\,=\,\zeta\,\mu_1$ and $\mu_1\,=\,.1$ (dashed line), $\mu_1\,=\,\zeta\,\mu_2$ and $\mu_2\,=\,.1$ (dotted line) \cite{FOG}. The behavior of GR is shown by the solid line. The dimensionless quantity $\zeta$ runs between $0\div 10$ with step $2$. The dimension of $\mu_1$ and $\mu_2$ is the inverse of length.}
  \label{plotpotential}
\end{figure}

Two last remarks on the (\ref{sol_gen}) are needed. The two corrections have different algebraic sign, and in particular the Yukawa correction with $\mu_1$ implies a stronger gravitational force, while the second one (purely induced by Ricci and Riemann square) contributes with a repulsive force. By remembering that the motivations of extending the outcome of GR to new theories is supported by missing matter justifying the flat rotation curves of galaxies, the first correction is a nice candidate. A crucial point is given by the spatial range of correction. In fact the Yukawa corrections imply a massive propagation; then, more massive is the particle, shorter is the spatial range. In Fig. \ref{plotpotential} we report the spatial behavior of gravitational potential (\ref{sol_pointlike}) for arbitrary values interval of parameters $\mu_1$ and $\mu_2$.

At last in Newtonian Mechanics the Gauss theorem gives us a spherically symmetric gravitational potential even if the spherically symmetric source is rotating. In GR as well as in FOG, however, the rotating spherically symmetric source generates an axially symmetric space-time (the well known Kerr metric) and only if the source is at rest one has the space-time with the same symmetry (the well known Schwarzschild metric). Then the galaxy being a rotating system will generate an axially symmetric space-time while we are using the solution (\ref{sol_gen}). This aspect is not contradictory because the solutions are calculated in the Newtonian limit (\emph{i.e.} $v^2\,\ll\,1$) and under this assumption the Kerr metric collapses into Schwarzschild metric. In fact we have

\begin{eqnarray}\label{Kerr-Scwh}
  g^{Kerr}_{\mu\nu}\,=\,\begin{pmatrix}
  1-\frac{r_g r}{\Sigma^2}& 0 & 0 & \frac{r_g r \eta}{\Sigma^2}\sin^2\theta\\
  \\
  0 & -\frac{\Sigma^2}{H^2} & 0 & 0\\
  \\
  0 & 0 & -\Sigma^2 & 0\\
  \\
  \frac{r_g r \eta}{\Sigma^2}\sin^2\theta & 0 & 0 & -\biggl(r^2+\eta^2+\frac{r_g r \eta^2}{\Sigma^2}\sin^2\theta\biggr)\sin^2\theta\\
  \end{pmatrix}\rightarrow
  \begin{pmatrix}
  1-\frac{r_g}{r}& 0 & 0 & 0\\
  \\
  0 & -1-\frac{r_g}{r} & 0 & 0\\
  \\
  0 & 0 & -r^2 & 0\\
  \\
  0 & 0 & 0 & -r^2\sin^2\theta\\
  \end{pmatrix}
\end{eqnarray}
where $r_g\,=\,2GM$ is the Schwarzschild radius, $\Sigma^2\,=\,r^2+\eta^2\cos^2\theta$, $H^2\,=\,r^2-r_g r+\eta^2$, $\eta\,=\,L/M$ and $L$ is the angular momentum along the z-axis.

\section{Mass model of galaxies}\label{mass_model}

From the point of view of morphology, a galaxy can be modeled by considering at least two
components: the bulge and the disk. Obviously the galaxy is a more complicated structure and there are others components, but for our aim this idealization is satisfactory. The bulge, generally, can be represented easily with cylindrical coordinates (but in a more crude idealization it is like a ball), while the disk has a radius bigger than the thickness. However we find that the rotation curve does not present the Keplerian behavior outside the matter, but the curve remain constant for any distance. Then we must formulate the existence of exotic matter that can justify the experimental observation. A simple discussion about the distribution of DM can be formulated by imposing the constant value of velocity in (\ref{circular_velocity}) for large distances. In fact we find

\begin{eqnarray}\label{density_DM}
v_c(r)\,\sim\,\text{constant}\,\rightarrow\,\rho_{DM}(r)\,\sim\,r^{-2}
\end{eqnarray}

A matter distribution as (\ref{density_DM}) has a problem when we want to calculate the total mass. In fact if we have $\rho\,\sim\,r^{-2}$, the mass diverges. A such exotic behavior seems no-physical, but this outcome is only the consequence of constant rotation curve. In fact by increasing the distance also the mass must increase with the power law for any distance (\ref{circular_velocity}). However, since the Gauss theorem holds in GR, the matter outside the sphere of integration does not contribute to the gravitational flux and then we do not have difference with respect to the ordinary matter.

This same argumentation is not valid in $f(X,Y,Z)$-Gravity: the no viability of Gauss theorem implies that the range of integration of DM could cover all range and also the matter outside is considered. A cut off is needed now. In this paper then for completeness we consider that the galaxy is composed by three components: the bulge, the disk and an alone of DM.

It should be noted that the spatial behaviors of DM (generally spherically symmetric) are made only \emph{a posteriori}: the cornerstones of study of rotation curves are the GR and the distribution of ordinary matter. Only after this assumption the distribution of DM is such as to justify the gap between the theoretical prediction and the experimental observation.

Before jumping to analysis of rotation curves we want to resume the principal spatial distributions of mass in the three galactic components. In literature there are many forms of density, but it is possible to resume them as follows.

More realistic models are the ones with mass density depending also on the $z$-coordinate for bulge and disk. Particularly one can consider the following choice

\begin{eqnarray}\label{density_1}
\left\{\begin{array}{ll}
\rho_{bulge}(R,z)\,=\,\frac{M_b}{4\pi\,V_0}\frac{{\xi_b}^{\gamma}}{[R^2+z^2/q^2]
^{\gamma/2}}\biggl[\frac{\xi_b+\sqrt{R^2+z^2/q^2}}
{\xi_b}\biggr]^{\gamma-\beta}\,e^{-\frac{R^2+z^2/q^2}{{\xi_t}^2}}\\\\
\rho_{disk}(R,z)\,=\,\frac{M_d}{4\pi\,{\xi_d}^2z_d}\,e^{-\frac{R}{\xi_d}
-\frac{|z|}{z_d}}\\\\
\rho_{DM}(r)\,=\,\frac{M_{DM}^{vir}}{4\pi{\xi_s}^3\,g(\xi_{DM}^{vir}/\xi_s)}\frac{\xi_s}{r}\frac{1}{[1+r/\xi_s]^2}
\end{array}\right.
\end{eqnarray}
In this choice, suggested by Dehnen \& Binney \cite{deh}, the bulge is described as a truncated power - law model (first line of
(\ref{density_1})) where $V_0\,=\,\int_{0}^{\infty}
dR'\,R'\,\int_{0}^{\infty}dz'\tilde{\rho}_{bulge}(R',z')$.
$\beta$, $\gamma$, $q$ and $\xi_t$ are the parameters. While for
the disk, one adopted a double exponential where the total
mass is $M_d\,=\,2\pi{\xi_d}^2\,\Sigma_\odot\,e^{\frac{\xi_0}{\xi_d}}$
with $\Sigma_\odot\,=\,(48\,\pm\,8)\,M_\odot/\text{pc}^2$ and
$\xi_0\,=\,8.5\,\text{Kpc}$ \cite{kui}. Finally in the case of DM the density profile is the NFW model \cite{navarro, NFW} where $g(x)\,=\,\ln(1+x)-\frac{x}{x+1}$, $M_{DM}^{vir}$ and
$\xi_{DM}^{vir}$ are the virial mass and virial radius and $\xi_s$
is a characteristic length. For the Milky Way
$\xi_{DM}^{vir}/\xi_s\,=\,10\,\div\,15$.

Leaving the axis-symmetry one can consider a more simple model:
the spherical symmetry model. With this approach one has \cite{wyse, noo, bin}

\begin{eqnarray}\label{density_2}
\left\{\begin{array}{ll}
\rho_{bulge}(r)\,=\,\frac{k\,M_b}{4\eta\,\pi\,{\xi_b}^{9/4}}\int_r^\infty dx\,\frac{e^{-k[(x/{\xi_b})^{1/4}-1]}}{x^{3/4}\sqrt{x^2-r^2}}\\\\
\sigma_{disk}(R)\,=\,\frac{M_d}{2\pi\,{\xi_d}^2}\,
e^{-\frac{R}{\xi_d}}\\\\
\rho_{DM}(r)\,=\,\frac{M_{DM}}{2(4-\pi)\pi{\xi_{DM}}^3}\,\frac{1}{1+\frac{r^2}{{\xi_{DM}}^2}}
\end{array}\right.
\end{eqnarray}
where $\xi_b$, $\xi_d$, $\xi_{DM}$, $M_b$, $M_d$, $M_{DM}$, are
the radii and the masses of bulge, disk and DM.
$k\,=\,7.6695$ and $\eta\,=\,22.665$ are dimensionless constants.
The density profile of bulge considered is the well-known
formula of de Vaucouleurs \cite{vau}.

A more simple model, resuming the previous ones, can be

\begin{eqnarray}\label{density_3}
\left\{\begin{array}{ll}
\rho_{bulge}(r)\,=\,\frac{M_b}{2\,\pi\,{\xi_b}^{3-\gamma}\,\Gamma(\frac{3-\gamma}{2})}\frac{e^{-\frac{r^2}{{\xi_b}^2}}}{r^\gamma}\\\\
\sigma_{disk}(R)\,=\,\frac{M_d}{2\pi\,{\xi_d}^2}\,
e^{-\frac{R}{\xi_d}}\\\\
\rho_{DM}(r)\,=\,\frac{\alpha\,M_{DM}}{\pi\,(4-\pi){\xi_{DM}}^3}\,\frac{1}{1+\frac{r^2}{{\xi_{DM}}^2}}
\end{array}\right.
\end{eqnarray}
where $\Gamma(x)$ is the Gamma function, $0\,\leq\,\gamma\,<\,3$
is a free parameter and $0\,\leq\,\alpha\,<\,1$ is the ratio of DM
inside the sphere with radius $\xi_{DM}$ with respect to the total DM. The radius $\xi_{DM}$ and the mass $M_{DM}$ play conceptually the same role respectively of $\xi_{DM}^{vir}$ and $M_{DM}^{vir}$. However, as before claimed, the hot point is the choice of DM model. Given all these models it is almost normal that there are many different estimates of DM. Therefore, the parameters of DM model may not be unique \cite{bur}.

\section{Rotation curves by $f(X,Y,Z)$-Gravity}\label{curve_FOG}

We are interesting to evaluate the circular velocity (\ref{stazionary_motion}) adopting the mass models (\ref{density_3}). So the potential (\ref{sol_gen}) becomes

\begin{eqnarray}\label{sol_gen_3}
\Phi(r,R,z)\,=&&G\,\biggl\{\frac{2\,M_b}{3\,{\xi_b}^{3-\gamma}\,\Gamma(\frac{3-\gamma}{2})}\frac{1}{r}\int_0^{\infty}
dr'\,r'^{1-\gamma}\,e^{-\frac{r'^2}{{\xi_b}^2}}\biggl[3\,\frac{|r-r'|-r-r'}{2}\nonumber
\\\nonumber\\\nonumber\\&-&\frac{e^{-\mu_1a|r-r'|}-e^{-\mu_1a(r+r')}}{2\,\mu_1\,a}
+2\,\frac{e^{-\mu_2a|r-r'|}-e^{-\mu_2a(r+r')} }{\mu_2\,a}
\biggr]\nonumber\\\nonumber\\\nonumber\\
&+&\frac{4\,\alpha\,M_{DM}}{3(4-\pi)\,{\xi_{DM}}^3}\frac{1}{r}\int_0^{\Xi/a}
dr'\,\frac{r'}{1+\frac{r'^2}{{\xi_{DM}}^2}}\biggl[3\,\frac{|r-r'|-r-r'}{2}\nonumber
\\\nonumber\\\nonumber\\&-&\frac{e^{-\mu_1a|r-r'|}-e^{-\mu_1a(r+r')}}{2\,\mu_1\,a}
+2\,\frac{e^{-\mu_2a|r-r'|}-e^{-\mu_2a(r+r')} }{\mu_2\,a}
\biggr]\nonumber\\\nonumber\\\nonumber\\
&-&\,\frac{M_d}{\pi\,{\xi_d}^2}\,\biggr[\int_0^\infty
dR'\,e^{-\frac{R'}{\xi_d}}\,R'\,\biggl(\frac{\mathfrak{K}(\frac{4RR'}{(R+R')^2+z^2})}{\sqrt{(R+R')^2+z^2}}+
\frac{\mathfrak{K}(\frac{-4RR'}{(R-R')^2+z^2})}{\sqrt{(R-R')^2+z^2}}\biggr)
\nonumber\\\nonumber\\\nonumber\\
&+&\int_0^\infty dR'\,e^{-\frac{R'}{\xi_d}}\,R'\,\int_0^{\pi}
d\theta'\frac{e^{-\mu_1a\,\Delta(R,R',z,0,\theta')}
-4\,e^{-\mu_2a\,\Delta(R,R',z,0,\theta')}}{3\,\Delta(R,R',z,0,\theta')}\biggr]\biggl\}
\end{eqnarray}
where $\Xi$ is the distance on which we observe the rotation
curve, $\mathfrak{K}$ is the elliptic function and the modulus of
distance is given by

\begin{eqnarray}\label{distance}
\Delta(R,R',z,z',\theta')\,\doteq\,|\mathbf{x}-\mathbf{x}'|\,=\,\sqrt{(R+R')^2+(z-z')^2-4RR'\cos^2\theta'}\
\end{eqnarray}
The constant $a$ is a scale factor defined by the
substitution $R\,, r\,\rightarrow\,a\,r\,,a\,R$ so all quantities are
dimensionless. At last, by remembering $r\,=\,\sqrt{R^2+z^2}$ the circular speed (\ref{stazionary_motion}) in the galactic plan is given by

\begin{eqnarray}\label{velocity}
v_c(R)\,=\,\sqrt{R\,\frac{\partial}{\partial R}\,\Phi(R,R,0)}
\end{eqnarray}

\begin{figure}[htbp]
  \centering
  \includegraphics[scale=1]{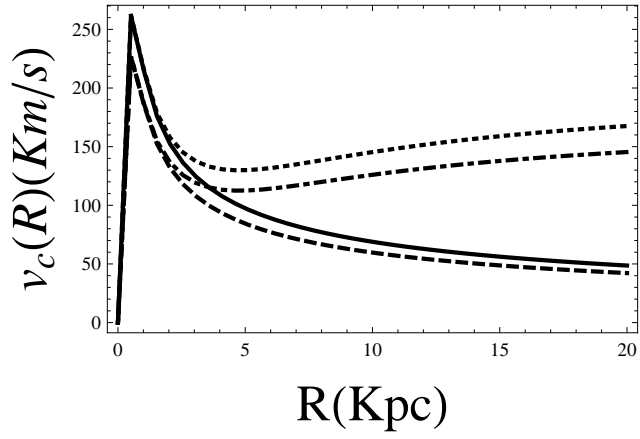}\\
  \caption{The rotation curve induced by bulge component (first line of (\ref{density_3})): GR (dashed line), GR + DM (dashed and dotted line), FOG (solid line) and FOG + DM (dotted line).}
  \label{vel_1}
\end{figure}

\begin{figure}[htbp]
  \centering
  \includegraphics[scale=1]{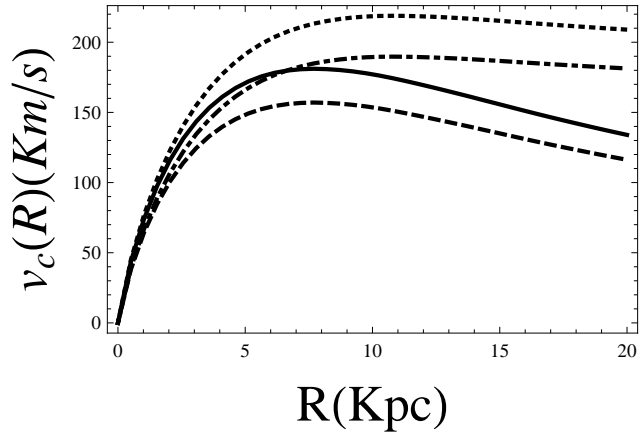}\\
  \caption{The rotation curve induced by disk component (second line of (\ref{density_3})): GR (dashed line), GR + DM (dashed and dotted line), FOG (solid line) and FOG + DM (dotted line).}
  \label{vel_2}
\end{figure}

In the Figs. \ref{vel_1}, \ref{vel_2}, we report the spatial behaviors of rotation curve induced by the bulge and disk component. The behavior for any component is compared in the framework of GR, FOG, GR + DM and FOG + DM. The values of free parameters of model are in the first line in Table \ref{tab_1} and refereing to Milky Way. The values of scale lengths $\mu_1$, $\mu_2$ are set at $10^{-2}\,a^{-1}$, $10^2\,a^{-1}$. In both components we note for $R\,>>\,\xi_b, \xi_d$ the Keplerian behavior, while it is missing only when we consider also the DM component. The shape of the rotation curve is similar to ones obtained by varying the total mass and scale radius. For a given scale radius, the peak velocity varies proportionally to a square root of the mass. For a fixed total mass, the peak-velocity position moves inversely proportionally to the scale radius, or along a Keplerian line.

\begin{figure}[htbp]
  \centering
  \includegraphics[scale=1]{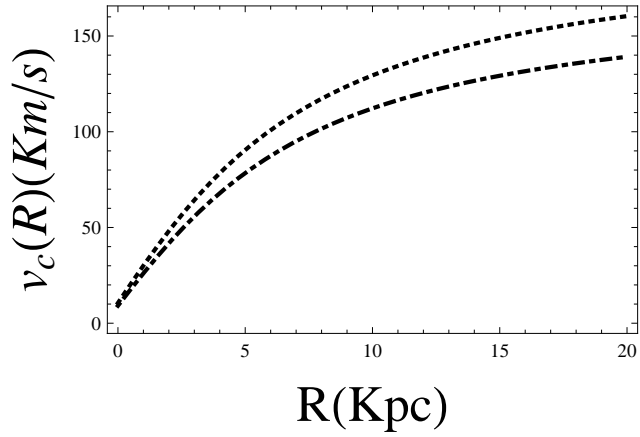}\\
  \caption{The rotation curve induced by DM component (third line of (\ref{density_3})): GR + DM (dashed and dotted line) and FOG + DM (dotted line).}
  \label{vel_3}
\end{figure}

As it is known in literature $f(X,Y,Z)$-Gravity, and in particular $f(X)$-Gravity, mimics a partial contribution of DM. In fact the corrective term $\propto\,e^{-\mu_1|\mathbf{x}|}/|\textbf{x}|$ contributes to enhance the attraction and thus the rotation curve must increase to balance the force. In the case of the other term, we have a correction $\propto\,-e^{-\mu_2|\mathbf{x}|}/|\textbf{x}|$ that contributes, being repulsive, to decrease the velocity. However in both cases these terms are asymptotically null and $f(X,Y,Z)$-Gravity and GR must lead to the same result. Only with the addition of DM it is possible to raise the curve and have almost constant values. In the Fig. \ref{vel_3} we report the component of rotation curve induced by only auto-gravitating DM.

\begin{table}[hbt]
    \begin{center}
    \caption{Parameters of models (\ref{density_3}). The unity of mass is $10^{10}\,M_\odot$ and $a\,=\,1\,\text{Kpc}$.}
    \begin{tabular}{lcccccccccc}
        \hline
        \hline
         Galaxy & $M_b$ & $\xi_b$ & $\gamma$ & $M_d$ & $\xi_d$ & $M_{DM}$ & $\xi_{DM}$ & $\alpha$ & $\Xi$ \\
        \hline
         Milky Way & 0.77 & 0.5 & 1.5 & 5.20 & 3.5 & 1.68 & 5.5 & 0.50 & 20 \\
         NGC 3198 & 0 & / & / & 2.60 & 3.5 & 0.84 & 5.5 & 0.53 & 20 \\
        \hline
        \hline
    \end{tabular}
\label{tab_1}
\end{center}
\end{table}

\begin{figure}[htbp]
  \centering
  \includegraphics[scale=1]{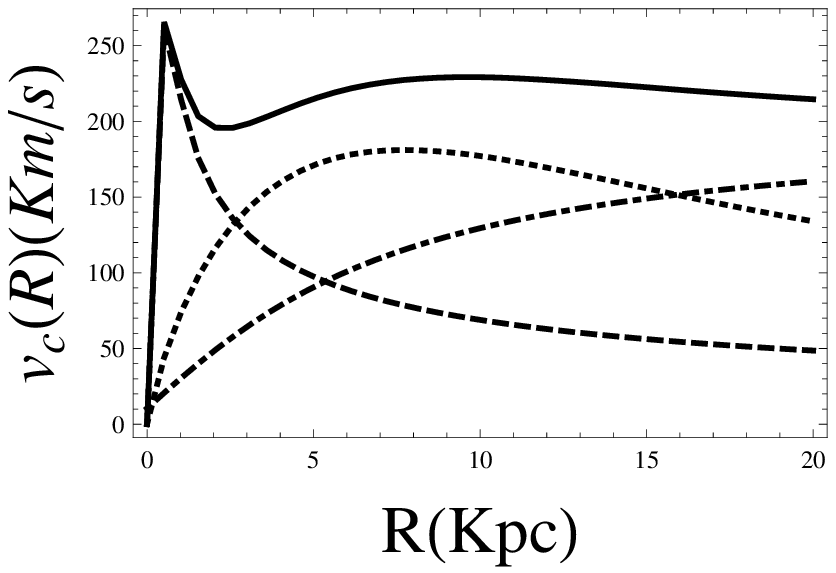}\\
  \caption{Comparison between the rotation curves of galactic components: bulge (dashed line), disk (dotted line), DM (dotted and dashed line) and the global galactic rotation curve (solid line). All curves have been valuated in the framework of FOG + DM.}
  \label{vel_4}
\end{figure}

\begin{figure}[htbp]
  \centering
  \includegraphics[scale=1]{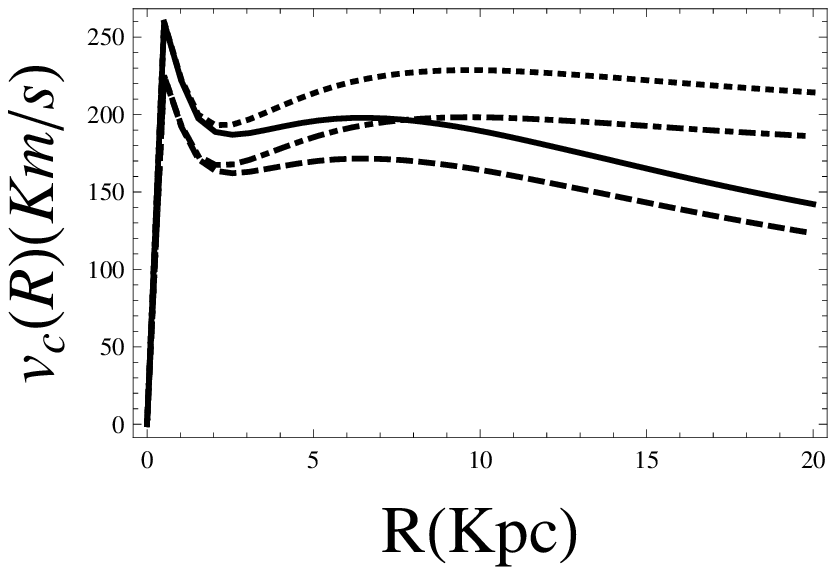}\\
   \caption{The global rotation curve in the framework of GR (dashed line), GR + DM (dashed and dotted line), FOG (solid line) and FOG + DM (dotted line). The values of "masses" are $\mu_1\,=\,10^{-2}\,a^{-1}$ and $\mu_2\,=\,10^2\,a^{-1}$.}
  \label{vel_5}
\end{figure}

\begin{figure}[htbp]
  \centering
  \includegraphics[scale=1]{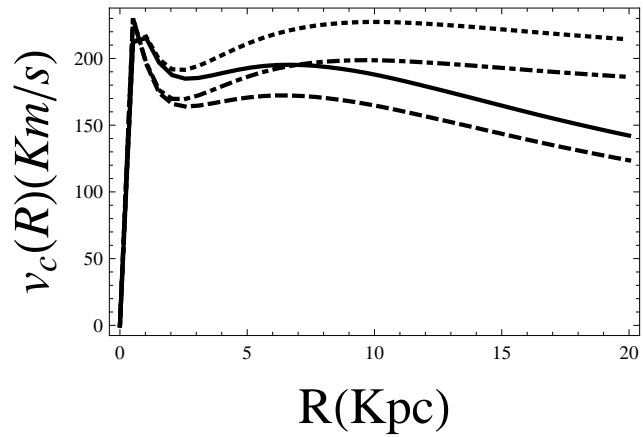}\\
   \caption{The global rotation curve in the framework of GR (dashed line), GR + DM (dashed and dotted line), FOG (solid line) and FOG + DM (dotted line). The values of "masses" are $\mu_1\,=\,10^{-2}\,a^{-1}$ and $\mu_2\,=\,5\,a^{-1}$.}
  \label{vel_6}
\end{figure}

In Fig. \ref{vel_4} we show the global behavior (experimentally expected) of rotation curve compared with respect to the bulge, disk and DM component for the $f(X,Y,Z)$-Gravity. While in Fig. \ref{vel_5} there is the global rotation curve in the framework of GR, FOG, GR + DM and FOG + DM. At last in Fig. \ref{vel_6} we replicate the outcome of Fig. \ref{vel_5} but we inserted the value $\mu_2\,=\,5\,a^{-1}$. In this case the rotation curve induced by $f(X,Y,Z)$-Gravity allows lower values as previously we claimed.


\begin{figure}[htbp]
  \centering
  \includegraphics[scale=1]{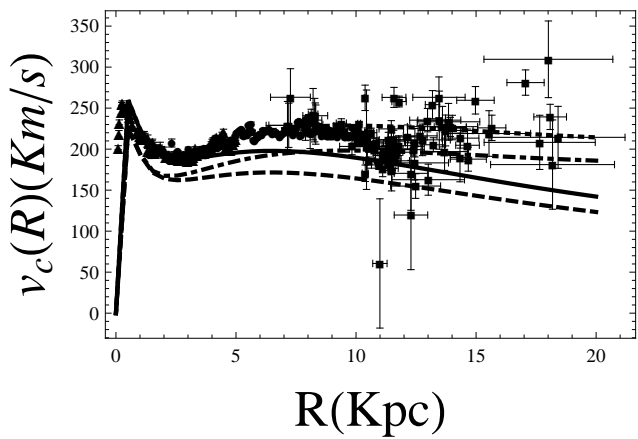}\\
  \caption{Superposition of theoretical behaviors (GR (dashed line), GR + DM (dashed and dotted line), FOG (solid line), FOG + DM (dotted line)) on the experimental data for Milky Way. The mass model used is shown in (\ref{density_3}) and the values of parameters are in Table \ref{tab_1}. The values of "masses" are $\mu_1\,=\,10^{-2}\,\text{Kpc}^{-1}$ and $\mu_2\,=\,10^2\,\text{Kpc}^{-1}$.}
  \label{vel_7}
\end{figure}

From the experimental point of view we used an updated rotation curve of Milky Way by integrating the existing data from the literature, and plot them in the same scale \cite{sofue}. The data used are available in a digitized from the URL http://www.ioa.s.u-tokyo.ac.jp/~sofue/mw/rc2009/. The unified rotation curve shows clearly the three dominant components: bulge, disk, and flat rotation due to the DM \cite{BG, CL, FBS, BFS, DB, sofue_2, sofue_3, HBCHI}. These data, finally, have been updated further by \cite{data_MW}. The whole set of data are plotted in Fig. \ref{vel_7} and on them the theoretical rotation curve induced by $f(X,Y,Z)$-Gravity with DM has been superimposed. The values of best fit are shown in Table \ref{tab_1} with $\mu_1\,=\,10^{-2}\,\text{Kpc}^{-1}$ and $\mu_2\,=\,10^2\,\text{Kpc}^{-1}$.

\begin{figure}[htbp]
  \centering
  \includegraphics[scale=1]{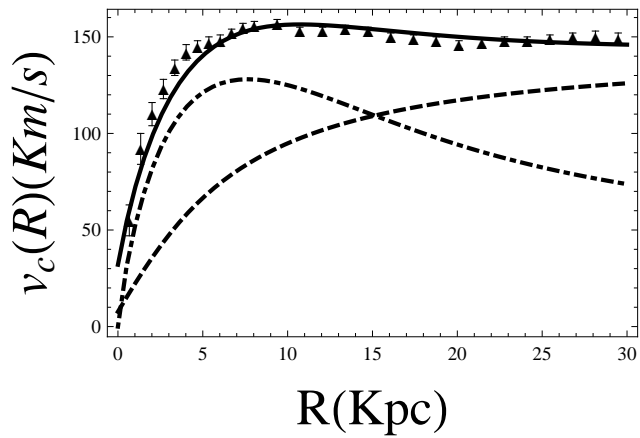}\\
  \caption{Superposition of theoretical behaviors (GR (dashed line), GR + DM (dashed and dotted line), FOG (solid line), FOG + DM (dotted line)) on the experimental data for NGC 3198. The mass model used is shown in (\ref{density_3}) and the values of parameters are in Table \ref{tab_1}. The values of "masses" are $\mu_1\,=\,10^{-2}\,\text{Kpc}^{-1}$ and $\mu_2\,=\,10^2\,\text{Kpc}^{-1}$.}
  \label{vel_8}
\end{figure}

The same mass model (\ref{density_3}) has been considered also for the galaxy NGC 3198. This galaxy has been chosen since the bulge is missing. Then we set $M_b\,=\,0$ in the (\ref{sol_gen_3}). In Fig. \ref{vel_8} we show the experimental data \cite{data_3198} and the superposition of theoretical behavior. Also in this case we find a nice outcome for a new set of parameters shown in Table \ref{tab_1}, while the values of $\mu_1$ and $\mu_2$ are the same of Milky Way.


By introducing the Gauss - Bonnet invariant defined by the relation $G_{GB}\,=\,X^2-4Y+Z$
\cite{dewitt_book} we obtain in four dimensions the condition  $H^{GB}_{\mu\nu}\,=\,H^{X^2}_{\mu\nu}-4H^{Y}_{\mu\nu}+H^{Z}_{\mu\nu}\,=\,0$. This condition on the field equations (\ref{fieldequationFOG}) enables us to consider only two curvature invariants \cite{FOG}, and by resolving the system (\ref{mass_definition}) with respect to the quantities $f_{XX}(0)$ and $f_Y(0)$ the general $f(X,Y,Z)$-Gravity can be recast as the effective Lagrangian (\emph{Quadratic Lagrangian}) in the Newtonian Limit

\begin{eqnarray}\label{FOG_theory_WF}
\mathcal{L}\,=\,f(X,Y,Z)\,=\,R-\frac{1}{3}\biggl[\frac{1}{2\,{\mu_1}^2}+\frac{1}{{\mu_2}^2}\biggr]\,R^2
+\frac{R_{\alpha\beta}R^{\alpha\beta}}{{\mu_2}^2}
\end{eqnarray}
The theory of Gravity represented by the Lagrangian (\ref{FOG_theory_WF}) is the more general theory considering all invariant curvatures, but we note a degeneracy. In fact we can have different $f(X,Y,Z)$-Gravity describing however the same Newtonian Limit \cite{newtonian_limit_fR, PRD, FOG}. The solution of field equations or the experimentally detectable quantities, as the rotation curve, are parameterized only by the derivatives of $f$, then we can have different functions $f(X,Y,Z)$ which admit the same physics.

The initial aim, \emph{i.e.} to extend the GR to a new class of theories, as we claimed in the introduction, is to justify the rotation curve without the DM component. From the previous outcomes, we see that even if the $f(X,Y,Z)$-Gravity, or better a $f(X)$-Gravity, admits a stronger attractive force, it is unable to realize our aim. \emph{Also in this framework we need Dark Matter}. Obviously we need a smaller amount of DM on the middle distances, but for large distances we have the same problems of GR.

\section{$R^n$-Gravity vs $f(X,Y,Z)$-Gravity}\label{CCT_theory}

The problem of DM seems to have been solved in literature, in the framework of $f(X)$-Gravity, by considering the Lagrangian $\mathcal{L}\,=\,R^n$ with $n\,\epsilon\,\mathrm{Q}$ \cite{CCCT, CCT}. In these papers the gravitational potential for a point-like source can be

\begin{eqnarray}\label{pot_CCCT}
\Phi_{R^n}(r)\,=\,-\frac{GM}{r}\biggl[1+\frac{(r/{r_c})^\beta-1}{2}\biggr]
\end{eqnarray}
where $r_c$ is a characteristic length and $\beta$ is a dimensionless parameter. To recover the condition $\lim_{r\rightarrow\infty}\Phi_{R^n}(r)\,=\,0$ one must have $0\,\leq\,\beta\,<\,1$. In the case $\beta\,=\,0$ the GR is found.

\begin{figure}[htbp]
  \centering
  \includegraphics[scale=1]{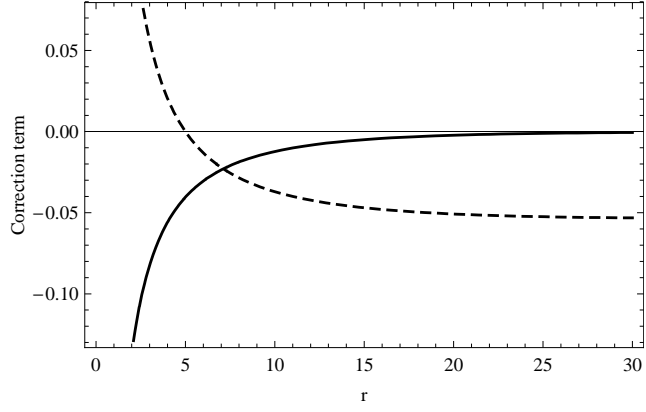}\\
  \caption{Comparison between the corrective terms induced by $f(X)$-Gravity ($-\frac{e^{-\mu_1\,r}}{3\,r}$, solid line) and $R^n$-Gravity ($\frac{1-(r/{r_c})^\beta}{2\,r}$, dashed line). $\mu_1\,=\,0.1$, $\beta\,=\,0.8$ and $r_c\,=\,5$. The unities for $r_c$ and $\mu_1$ are arbitrary. The dashed curve shows a very slow ascent.}
  \label{Y_CCCT}
\end{figure}

We comment about the physical behavior of potential (\ref{pot_CCCT}) and we want to add some reflections considering the result of the rotation curve shown above. Before to analyze the mathematical properties of metric linked to potential (\ref{pot_CCCT}), we want to show the different values of correction to the Newtonian potential. In Fig. \ref{Y_CCCT} we report the radial behavior of the corrections to $1/r$ for the potentials (\ref{sol_pointlike}) and (\ref{pot_CCCT}) (to minimize the difference we considered only $f(X)$-Gravity). From the plot we note a discrepancy between the two corrections. The correction by $\biggl(R-\frac{R^2}{6\,{\mu_1}^2}\biggr)$-Gravity acts over distances much smaller, while the correction induced by $R^n$-Gravity provides a potential nearly constant over large intervals and slowly goes to zero ($\sim\,r^{\beta-1}$). For this aspect the potential (\ref{pot_CCCT}) does not need the DM component. Then with a procedure of fine tuning of $r_c$ and $\beta$ it was possible to justify the experimental rotation curve for a wide class of galaxies \cite{CCT} when $n\,=\,3.5$. This choice was possible because there must be a relationship $\beta\,=\,\beta(n)$ so that the potential (\ref{pot_CCCT}) was compatible with respect to the field equations. These are the positive aspects of the potential (\ref{pot_CCCT}) used in \cite{CCCT, CCT}.

We conclude this section by reviewing the fundamental weaknesses of $R^n$-Gravity. \textbf{A}) The potential (\ref{pot_CCCT}) presents an analogous behavior of potential (\ref{sol_pointlike}). In fact for $r\,<\,r_c$ one has $\frac{1-(r/{r_c})^\beta}{2\,r}\,>\,0$ then the correction is "repulsive" like one induced by Ricci tensor square, while for $r\,>\,r_c$ one has an attractive correction. Now by remembering the reason of extension of GR, now we have unlike a repulsive contribution for $r\,<\,r_c$. If in $f(X,Y,Z)$-Gravity we can delete the Ricci tensor square contribution and we have only the $f(X)$-Gravity, in $R^n$-Gravity we must collapse only in GR. \textbf{B}) The potential (\ref{pot_CCCT}) belongs to general class of solutions for $R^n$-Gravity classified by a perturbative method \cite{spher_sym}, but the solutions are $n$-independent. Obviously the general solutions (it would be hard challenge to find them) are $n$-dependent, but at first order with respect to the perturbative parameter and in the vacuum \footnote{This parameter is generally $c^{-2}$, but the analysis is the same if we consider the dimensionless ratio $r_g/r$.} the field equations are identically vanishing. So we say that the presence of matter has been not considered and the choice of arbitrary constant has been evaluated only by matching $R^n$-Gravity with GR in the limit $\beta\,\rightarrow\,0$. In fact by solving the field equations correctly in presence of matter (also with the point-like source) we would obtain solutions depending on the perturbative parameter and the technique is misplaced.

For these two aspects, but especially for the point \textbf{B}), $R^n$-Gravity does not admit the Newtonian limit if $n\,\neq\,1$. The potential (\ref{pot_CCCT}) does not follow a correct framework when extending the GR to the new theories we want to generalize the Newtonian potential. Generally all theories without Ricci scalar in the Lagrangian suffer from the same problem. For example also $R^2$-Gravity is in the same situation: is not possible to extend the solution in the matter \cite{Stabile_Capozziello}. \emph{Although we have solutions as $1/r$ with additional asymptotically flat terms, is not automatic the assertion that these solutions are the Newtonian limit of theory}.

Let us analyze now the mathematical properties of the metric trying to justify the difference of spatial behaviors in Fig. \ref{Y_CCCT}. To simplify the calculation we choose a set of standard coordinates. The metric (\ref{metric_tensor_PPN}), from the expressions (\ref{sol_pointlike}), becomes\footnote{The set of standard coordinates is defined by the condition to obtain the standard definition of the circumference with radius $r$. From the metric tensor (\ref{metric_tensor_PPN}) we must impose the condition $\biggl[1-2\Psi(r)\biggr]r^2\,=\,\tilde{r}^2$ for the new radial coordinate.}

\begin{eqnarray}\label{standard_metric}
  ds^2\,=\,\biggl[1-\frac{r_g}{r}\biggl(1+\frac{1}{3}\,e^{-\mu_1r}
  -\frac{4}{3}\,e^{-\mu_2r}\biggr)\biggr]dt^2-
  \biggl[1+\frac{r_g}{r}\biggl(1-\frac{\mu_1r+1}{3}\,e^{-\mu_1r}
  -\frac{2(\mu_2r+1)}{3}\,e^{-\mu_2r}\biggr)\biggr]dr^2-r^2d\Omega
\end{eqnarray}
where $d\Omega\,=\,d\theta^2+\sin^2\theta d\phi^2$ is the solid
angle, while the element of distance linked to potential (\ref{pot_CCCT}) can be written as follows

\begin{eqnarray}\label{me_stand}
ds^2\,=\,\biggl[1+2\Phi^{SC}_{R^n}(r)\biggr]\,dt^2-\biggl[1-2\Psi^{SC}_{R^n}(r)\biggr]\,dr^2-r^2d\Omega
\end{eqnarray}
where $\Phi^{SC}_{R^n}(r)$ is the the potential (\ref{pot_CCCT}) and $\Psi^{SC}_{R^n}(r)$ is the other potential missing in the paper \cite{CCT}. However in their analysis the knowledge of last potential is useless because its contribution in the geodesic motion is at fourth order. By following the paradigm of weak field limit at small velocity \cite{newtonian_limit_fR, PRD} for the $R^n$-Gravity we find

\begin{eqnarray}\label{me}
\Psi^{SC}_{R^n}(r)\,=\,-\frac{GM+K(\beta)+K_X}{r}+\frac{\beta-1}{4}\frac{GM}{r}\biggl(\frac{r}{r_c}\biggr)^\beta+
\frac{1}{4\,r}\int dr\,r^2 X(r)
\end{eqnarray}
where $K(\beta)$ and $K_X$ are constants depending, respectively, on the value of $\beta$ and on the integral operation, while the Ricci scalar $X$ could be an arbitrary function. In fact it needs some comment about the index $n$ in the Ricci scalar. If $n$ is a integer number, then the Ricci scalar can assume any value and can be also a generic space depending function. More attention is needed if $n$ is a rational number. The field equations (\ref{fieldequationFOG}) take into account up to third derivatives with respect to the Ricci scalar, then we must ensure that the function $f(X)$ and its derivatives are always well defined \cite{spher_sym}. In this case for $n\,<\,3$ the solution Ricci flat ($X\,=\,0$) or space depending and asymptotically vanishing are excluded. Only solutions with constant values are allowed, but the algebraic sign is crucial. A such behavior is expected any time we have the condition $\lim_{X\rightarrow 0}\,f(X)\,=\,\text{constant}$ \cite{spher_sym}. Now in all these considerations we do not recover the condition $\lim_{r\rightarrow\infty}\Psi^{SC}_{R^n}(r)\,=\,0$: then we have a theory which do not provide us the Minkowskian limit. It is using the first perturbative contribution of a metric component (providing the flatness at infinity), while other contributions in the remaining metric components (negligible in the Newtonian limit) do not cover the same asymptotic limit. Then only for $n\,>\,3$ we can have the flatness at infinity.

Then we could say that for $n\,>\,3$ the Minkowskian limit is recovered but a such perturbative approach can be performed only in the vacuum. The objection previously shown comes back. Up to third order ($c^{-6}$ or ${r_g}^3$) the geometrical side of field equation is identically null, but the matter side could not be null at first order ($c^{-2}$ or $r_g$).

The class of $R^n$-Gravity are examples of theories where the weak field limit procedure does not generate automatically the Minkowskian limit. In fact only if we consider theories satisfying the condition $\lim_{X\rightarrow 0}\,f(X)\,=\,0$ \cite{spher_sym}, their weak field limit is compatible with the request of asymptotically flatness. Moreover $f(X)$-Gravity mimicking an additional source due to its scalar curvature \cite{newtonian_limit_fR, stab_PM} we would have a constant matter that pervades all space giving us a justification of more intense gravitational potential. In addition if $\lim_{X\rightarrow 0}\,f(X)\,=\,\text{costant}$ we do not have the Minkowskian limit, but we can interpret the apparent mass, only from the experimental point of view, as DM. These aspects, then, can be a mathematical motivation for different shape of point-like gravitational potential, but also source of further attention.

\section{Conclusions}\label{conclusion}

In this paper we computed the study of galactic rotation curve when a FOG is considered. Among the several theories of fourth order we considered a generic function of Ricci scalar, Ricci and Riemann tensor. We started from the outcome of previous papers about the point-like solutions in the so-called weak field limit of theory and formulated the expression of rotation curve by inserting in the model the principal galactic components: the bulge, the disk and the DM component. In this limit we used the superposition principle for the potential since the Newtonian limit correspond also to the linearized version of theory. Since in FOG the Gauss theorem is not valid, the calculus of potential is performed directly by integrating in all space and obtaining in the potential also the information about the spatial shape of mass distribution. Apart the mathematical difficulties incoming from the research of gravitational potential the non-validity of Gauss theorem implies, for example, that a sphere can not be generally reduced to a point.

Among two corrective contributions Yukawa-like to Newtonian potential seems that only one induced by a generic function of Ricci scalar is a good candidate; in fact this contribution has the same algebraic sign of Newtonian component and we can detect a more attractive force. A such situation can justify partially the bigger observed velocity, while the term induced by Ricci tensor (and by the Riemann tensor) acts lowering the velocity.

The rotation curves have been evaluated by considering the bulge and the DM component spherically symmetric and the disk as a circular plane where the radius is larger than the thickness. Also in our case of FOG with ordinary matter we find that the rotation curve has the Keplerian behavior and only if we add the DM component we have a nice matching between theoretical and experimental data. However the hypothesis of existence of DM make two serious problems: since the matter distribution of DM is diverging when we consider the whole amount of mass is crucial the choice of cut-off inside the integral. Other hot point is the choice of the mass model of DM, or minimally the values of free parameters in the model. If we consider the GR as the theory of Gravity, we can have different values from those obtained if we choice the FOG. The spatial behaviors of DM (generally spherically symmetric) are made only \emph{a posteriori}.

The proposed models have been compared with the data of Milky Way and NGC 3198 obtaining the best fit for $\mu_1\,=\,10^{-2}\,\text{Kpc}^{-1}$ and $\mu_2\,=\,10^2\,\text{Kpc}^{-1}$. In both cases we found a nice outcome for given values of free parameters of mass model. However the initial aim to extend the GR to a new class of theories, as we claimed in the introduction, has the hope to justify the \emph{rotation curve without the DM component}. From the previous outcomes, while the $f(X,Y,Z)$-Gravity, or better a $f(X)$-Gravity, admits a more attractive force, on the other hand these modifications are not successful. \emph{Also in this framework we need Dark Matter}. Obviously we need a smaller amount of DM on the middle distances, but for large distances we have the same problems of GR.

We conclude the paper by comparing our outcome with the one of $R^n$-Gravity. In fact the problem of DM seemed to be solved by using a such theory. Particularly the constant rotation curves would be obtained without the DM. By finding a corrective term as power law in the radial distance one had a nice matching. But the model is not completely satisfactory. It is in our opinion that $R^n$-Gravity is misplaced but it contains some interesting and valid contributions. If the comparison with the experimental data encourages us to continue on this road, on the other hand, there are some mathematical pathologies that must be carefully evaluated.

Furthermore, we think that a good weak field limit at small velocity, \emph{i.e.} the Newtonian limit, of any theory of Gravity must always satisfy the condition of Minkowskian flatness. In addition, a nice self-consistent theory, from a mathematical point of view, must be placed. It is obvious that can be considered also the weak field limit on the cosmological background and in this case we do not require the Minkowskian limit.

Since, today, there is not a successful model to explain the rotation curve without requiring the existence of DM, but there are many partially acceptable models, we can say that, regardless of its detection, it remains an hard challenge to interpret the Dark Matter as a single geometric phenomenon.

\section{Acknowledges}

The authors would like to thank V.F. Cardone and Y. Sofue for their useful suggestions about the galactic experimental data and F. Siano for his advice on writing the software for numerical simulations.

\end{document}